# Variation in interface strength of Silicon with surface engineered $Ti_3C_2$ MXenes

Vidushi Sharma[*1], Dibakar Datta[*1]

[1]Department of Mechanical and Industrial Engineering, New Jersey Institute of Technology, Newark, NJ 07103, USA

## Abstract

Current advancements in battery technologies require electrodes to combine high-performance active material such as Silicon (Si) with two-dimensional materials such as transition metal carbides (MXenes) for prolonged cycle stability and enhanced electrochemical performance. More so, it is the interface between these materials, which is the nexus for their applicatory success. Herein, the interface strength variations between amorphous Si and $Ti_3C_2T_x$ MXene are determined as the MXene surface functional groups ($T_x$) are changed using first principle calculations. Si is interfaced with three $Ti_3C_2$ MXene substrates having surface -OH, -OH and -O mixed, and -F functional groups. Density functional theory (DFT) results reveal that completely hydroxylated $Ti_3C_2$ has the highest interface strength of 0.6 $J/m^2$ with amorphous Si. This interface strength value drops as the proportion of surface -O and -F groups increases. Additional analysis of electron redistribution and charge separation across the interface is provided for a complete understanding of underlying physico-chemical factors affecting the surface chemistry and resultant interface strength values. The presented comprehensive analysis of the interface aims to develop sophisticated MXene based electrodes by their targeted surface engineering.

## 1. Introduction

The promise of silicon (Si) as a commercial anode for Li-ion batteries(LIB), premised by its very high specific capacity(~3000 $mAhg^{-1}$ A), has been in question for decades due to slow kinetics and stress-mediated mechanical failures[1]. To combat mechanical failures, additives are added to Si that acts as a mesh in the electrode architecture, providing Si enough space to expand and contract. Though initially polymers were used along with Si nanoparticles to create a breathable supportive mesh for long-lasting mechanical stability[2-4], they lacked the most essentially required ionic and electronic conductivity besides being the cause of the electrode system's added weight. Driven by the need to replace polymer additives with a conductive and flexible binder, 2D transition metal carbides/nitrides (MXenes), which were discovered by Gogotsi and coworkers in 2011[5] have been recently mixed with Si anode by diverse synthetic procedures[6-10]. MXenes are novel type of 2D structures that have gained the popular vote due to their high electronic conductivities, stability, hydrophilicity, and surface chemistry[11-15]. With these properties, and backed by their potential to have surface engineered by modulating the functional groups, MXenes promise excellent performance as electrodes and supercapacitors for Li and beyond batteries[16-20].

Experimental reports have shown that Si/MXene composite excels in performance over its parent Si anode in capacity retention and cycle stability[7, 9, 10]. Conductive MXene functions as more than a binder in the electrode system by providing additional diffusive pathways, enhancing electron transport, and acting as a current collector[19, 21]. Above all, it is the stability of Si and MXene's interface, which is the foundation for the Si/MXenes system's aforementioned efficacies. Si is previously known to suffer from mechanical strains during battery performance if interfaced firmly with a substrate (in generic sense additive, binder, or current collector)[22]. Therefore, interface adhesion of Si with substrate MXene needs to be critically tailored for optimum performance in batteries. MXenes have an added advantage of a vast library of materials (from which ~30 have been synthesized)[11], and the option of modulating surface terminations (-OH, -O, and -F) by the choice of exfoliating agent during the synthesis process[23]. These surface terminations have a compelling role in altering the surface properties of an MXene[24-26]. With these revelations, the research community quickly predicted the impact of surface functional groups on MXene's performance in LIB battery systems. Computational techniques have turned out to be the preferred mode of investigation to study their atomic-scale dynamics. Diffusion studies of different ions in the interlayer spaces of functionalized MXenes have indicated that the OH and F groups tend to form clusters with Li and provide steric hindrance during the diffusion process[27]. In contrast, O-functionalized MXenes have manifested improved electrochemical performance and larger LIB capacities[17]. Recently, some extrinsic functional groups were successfully incorporated on MXenes to enlarge the interlayer spaces for enhanced charge-discharge kinetics and improved energy storage[20, 28-30]. All attempts were targeted to alleviate the role of MXenes as an electrode, with no attention being given to exploring characteristics of the functional group driven interface of MXenes with Si.

As researchers advance towards utilizing surface terminations to influence electrode performance, there is a necessity to establish their impact on the interface adhesion strength between MXene and the bulk material such as Si. In recent years, adhesive interaction between electrode components has impacted the over-all electrode morphology, cycle life, and electronic performance[22, 31]. At present, most studies focus on the adhesion of a single atom[20, 32], or at most small atom clusters such as of Li-S on the MXene surface[33]. First principle calculations have indicated a linear correlation between adsorption energies of single transition metal atoms on $Ti_3C_2$ (earliest reported MXene, which also remains the most studied due to its superior conductivity) and chemical attributes, such as charge distribution, bond length, and d-electron center of metal[32]. Interface adhesion analysis between a MXene and a 3D bulk go up to a recent experimentally measured value of 0.90 J/m$^2$ between $SiO_2$ and $Ti_3C_2T_x$[34]. The study reports variation in interface adhesion between the two materials as atomic thickness of MXene monolayer is changed. In their experimental work, $Ti_3C_2T_x$ has higher adhesion with $SiO_2$ ( 0.9 J/m$^2$) which drops to 0.4 J/m$^2$ for $Ti_2CT_x$. By far, no focus has been laid on the specificity of MXene surface functional groups ($T_x$) in all these reports.

Thus, in the present study, we investigate the interface strength between 3D Si bulk and $Ti_3C_2T_x$ MXene with differing surface functionalities by means of first principle calculations. Amorphous Si is interfaced with -OH, -OH/O mixed, and -F functionalized $Ti_3C_2$ MXene at near uniform interfacial distances. Surface energies calculated using density functional theory (DFT) permit the determination of interface strength as work of separation ($W_{sep}$). The present investigation details the variation of Si/$Ti_3C_2T_x$ interface strength primarily with the changing surface functional group ($T_x$). Further, a comprehensive analysis of interfacial gap, surface chemistry, and electron redistribution at the Si/$Ti_3C_2T_x$ interface is done to describe better the physico-chemical phenomenon impacting the interface strength.

## 2. Models and computational details

Three $Ti_3C_2T_x$ MXenes with different surface functional groups ($T_x$) were modeled prior to the interface analysis. $Ti_3C_2T_x$ were derived from a stable and experimentally recognized atomic model of free-standing $Ti_3C_2$, where three Ti atomic layers are inter-cleaved with two C layers to result in five atomic thick $Ti_3C_2$ monolayer. Functional groups were attached to the surface under-coordinated Ti atoms, above the hollow site between three neighboring C atoms. Among all possible configurations of functional groups, this has been validated as thermodynamically most stable [35, 36]. The three $Ti_3C_2T_x$ configurations considered for the study are : *(i)* hydroxylated MXene $Ti_3C_2(OH)_2$, where surface is saturated with -OH functional group; *(ii)* mixed functionalized MXene $Ti_3C_2(OH/O)_2$ , where ~38% surface -OH groups are randomly replaced by -O; and *(iii)* fluorinated MXene $Ti_3C_2F_2$, having -F as the only surface functional group. The three starting models were used for further analysis after complete optimization using density functional theory (DFT) within Vienna Ab initio Simulation Package (VASP)[37]. The top view of three MXene configurations can be seen in figure 1(a1-c1).

Investigation of interface strength required surface energies of three MXene models, amorphous Si (*a*-Si) bulk, and the interface energy of *a*-Si/$Ti_3C_2T_x$ systems. Amorphous Si bulk having 64 Si atoms has been derived from a crystalline $Si_{64}$ (Diamond FCC) using the computational quenching process in accordance with our previous work[22]. Slabs of three MXenes and optimized *a*-Si were subjected to DFT relaxation with added an vacuum of 20 Å to calculate of surface energies. It was critical for *a*-Si's free surface in the vacuum slab model to have the same surface area as its substrate MXene. Since the surface area of three $Ti_3C_2$ MXene models differs slightly due to different surface functionalization, we used three different vacuum models for *a*-Si surface energy calculation, each corresponding to individual MXene configuration. For the interface models, three optimized MXenes were individually interfaced (as depicted in figure 1) with a relaxed structure of *a*-Si bulk at an initial interfacial gap (*d*) of ~2.3Å. The interfacial gap *d* throughout the study is considered to be the vertical distance between lowest Si atom and top surface atoms of the MXene substrate. This consideration of initial *d* for the interface strength calculation is based on two assumptions. The distance of 2-2.5 Å between current two material surfaces should be ideal to encourage bonding. Moreover, as the interface's interaction is sensitive to an interfacial gap, huge variation in interfacial gap among the three interface systems might not provide us actual impact of the surface chemistry on the interface for comparison. Next, the interface energies were calculated using a vacuum interface model[38] with an added vacuum of 20 Å in z-dimensions (normal to the free surface) to permit complete ionic relaxation and circumvent periodic images' influence.

All optimizations were done using DFT within the VASP package[37]. Projector-augmented-wave (PAW) potentials were used to mimic the inert core electrons and valence electrons were represented by plane-wave basis set with the energy cutoff of 650 eV[39, 40]. Conjugate gradient method was employed for energy minimization with Hellmann-Feynman forces less than 0.02 eV/Å and convergence tolerance set at $1.0\times10^{-6}$ eV. The GGA with the PBE exchange-correlation function was taken into account[41]. For all DFT calculations, gamma-centered 4 X 4 X 1 k-meshes were taken, and GGA functional was inclusive of vdW correction to incorporate the effect of weak long-range van der Waals (vdW) forces[42]. All calculations were done with optPBE functional within vdW-DF-family[43, 44]. In addition, three ab-initio molecular dynamics (AIMD) simulations were performed within the DFT framework of VASP. These simulations targeted to observe changes in the interface system energy as Si atoms of amorphous Si bulk diffuse in the interface region, causing variations in the interfacial gap $d$. AIMD simulations were run with 1 fs time interval and temperature set to 300 K within NVT ensemble. Plane-wave basis cutoff was set to 550 eV for AIMD, and 2 X 2 X 1 gamma centered k-meshes were taken into account.

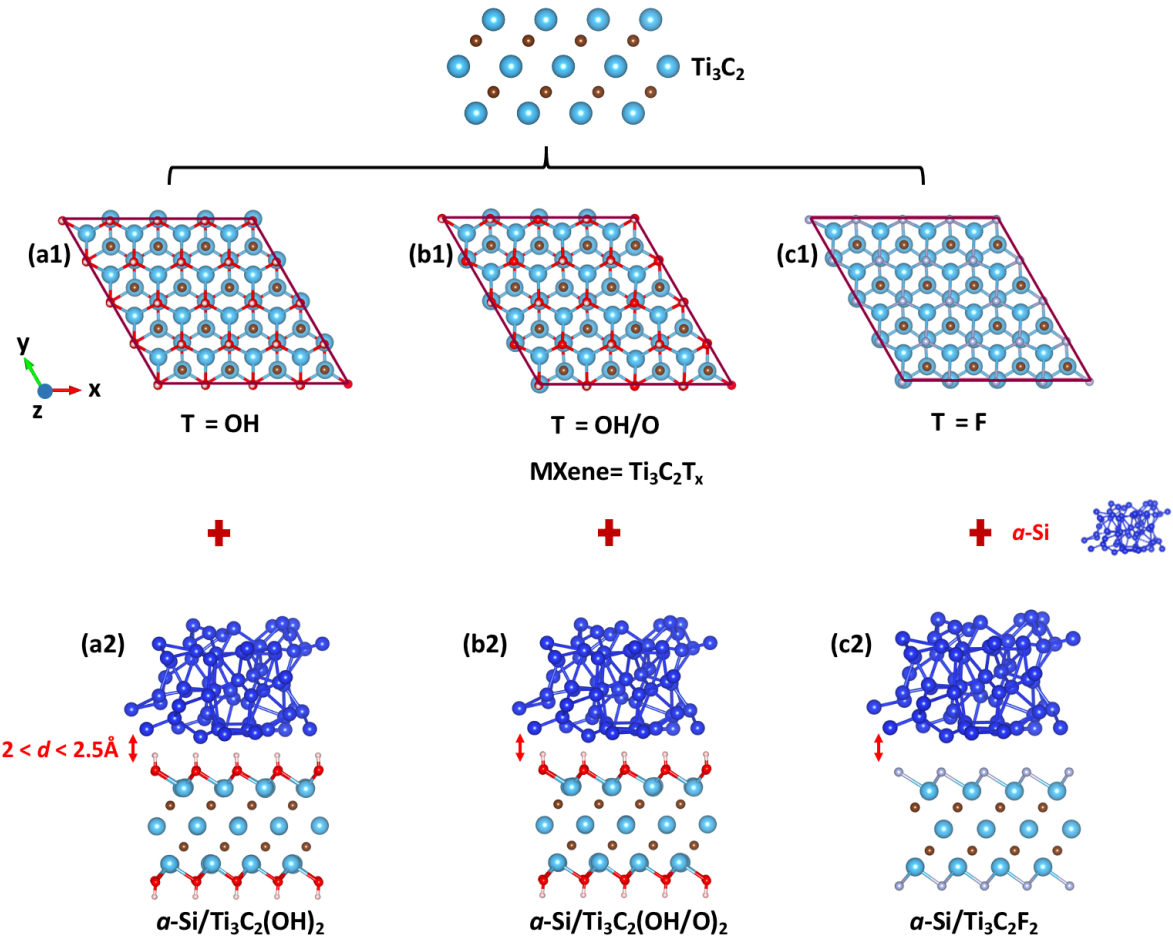

**Figure 1 Atomic representation of three $Ti_3C_2T_x$ MXenes and their initial Si/MXene interfaces.** **(a1,b1,c1)** Top view of surface functionalized $Ti_3C_2T_x$ MXene monolayers after DFT optimization. The surface functional groups ($T_x$) are changed from OH, to a combination surface of OH and O groups, and lastly, F. **(a2,b2,c2)** Optimized MXenes are interfaced with a relaxed amorphous Si (*a*-Si) at an interfacial gap *d* ranging from 2-2.5 Å for the interface energy calculations. The interfacial gap *d* is the vertical distance between lowest Si atom and top surface atoms of the MXene substrate.

## 3. Results and Discussion

MXenes are derived from bulk MAX phases via chemical exfoliation using hydrogen fluoride (HF). During experimental synthesis, prominent surface terminations are -OH and -F depending upon aqueous HF concentration used in the exfoliation process. Post chemical treatment, MXene is dried to remove the excess water, which can sometimes lead to cleaving of H from -OH surface terminations, resulting in $H_2$ release. This process leaves behind -O surface terminations. For the conversion of -OH termination to -O, additional energy of about 1.6 eV is required, and therefore, -O surface terminations are usually fewer in count[45, 46]. In most experimental synthesis, the MXene surface comprises of a mix of -OH, -O and -F groups. Still, surface functional groups can be carefully tailored by optimizing HF concentration and drying temperature during synthesis procedures. These devised surface groups can drastically change MXenes surface properties and interfere in interface attributes. Thus, in the following sections, we discuss the influence of changing functional groups on the strength of the interface between $Ti_3C_2$ MXenes and *a*-Si.

*3.1 Interface strength*

In order to obtain the interface strength of functionalized $Ti_3C_2T_x$ MXenes with Si, we created vacuum slab models for all three interface systems, as represented in figure 2a. Here, slab 1 consists of *a*-Si, slab 2 consists of functionalized $Ti_3C_2T_x$ MXene, and slab 3 have interface system of *a*-Si over the respective MXene. These structures are periodic in x-y dimensions with a vacuum of 20 Å in z dimension. Final energy outputs from the DFT simulations of slab models are listed in Table 1 and were used to calculate work of separation (**$W_{sep}$**). $W_{sep}$ is the energy required to completely separate the two materials at the interface, in a direction normal to the surface. The standard definition of $W_{sep}$ is:

$$W_{sep} = \sigma_1 + \sigma_2 - \gamma_{12} = \frac{E_1 + E_2 - E_{12}}{A} \tag{1}$$

Here, **$\sigma_1$, $\sigma_2$** are the surface energy of both materials in the system and **$\gamma_{12}$** is the interface energy[47]. These are determined from total energies of slab 1, slab 2 and slab 3 as **$E_1$, $E_2$** and **$E_{12}$**, respectively. **A** is the area of contact at the interface (surface area in x-y plane). Table 1 summarizes the slab energies $E_1$, $E_2$, $E_{12}$ and surface area A for all the three interface systems post optimization. The calculation of surface area for the individual systems is detailed in Supporting Information. To draw out distinctiveness in the interfacial interaction between *a*-Si and $Ti_3C_2T_x$ as $T_x$ is varied, it was important to maintain uniformity in the interfacial gap between *a*-Si and MXenes (*d*). During optimization of interface systems, Si atoms of amorphous bulk dispersed to their lowest energy positions resulting in a variable interfacial gap *d* between MXene and *a*-Si. Yet, final *d* remained between 2- 2.5Å as briefed in Table 1.

The interface strength result via $W_{sep}$ are presented in figure 2b and explicitly show that interface strength between *a*-Si and $Ti_3C_2T_x$ MXene change as the functional groups on MXene surface ($T_x$) are altered. The interface strength of *a*-Si/$Ti_3C_2(OH)_2$ is calculated to be 0.606 J/m² in $W_{sep}$ calculations. This presently derived interface strength is comparable in magnitude to the recent experimental results of $SiO_2$/$Ti_3C_2T_x$ (0.9 J/m²) and $SiO_2$/$Ti_2CT_x$ (0.4 J/m²) interfaces[34]. However, surface chemistry's role on adhesion interactions at the interface becomes more prominent as variation in the interface strength is seen even with the slightest change of surface functional groups on MXenes. The value of $W_{sep}$ dropped to 0.142 J/m² as fewer -OH groups are replaced by -O in the second interface system. Only 38% variation of the surface functional group (-OH to -O) significantly weakened the interface. With complete surface fluorination of $Ti_3C_2$, $W_{sep}$ value further dropped to 0.115 J/m².

The interface strength of *a*-Si with $Ti_3C_2T_x$ MXenes is below 0.6 J/m², alike the interface strength results presented by Basu et al.[22] between *a*-Si and graphene (0.41 J/m²) using the same methodology. This justifies why MXenes are increasingly being used along with Si in the battery systems. Interface adhesion of similar magnitude between active electrode particle and substrate benefits cycle life of a battery by mitigating stresses during lithiation/delithiation associated volume changes[22]. MXenes have proven to act as a promising substrate for active electrode particles such as Si, by effectively accommodating volume expansions and imparting system with flexibility for the generation of flexible stable electronics[7, 10]. Low interface strength between the system components is desired for liberal twisting and bending of MXenes, and prevent brittle failures associated with strong interfacial bonding.

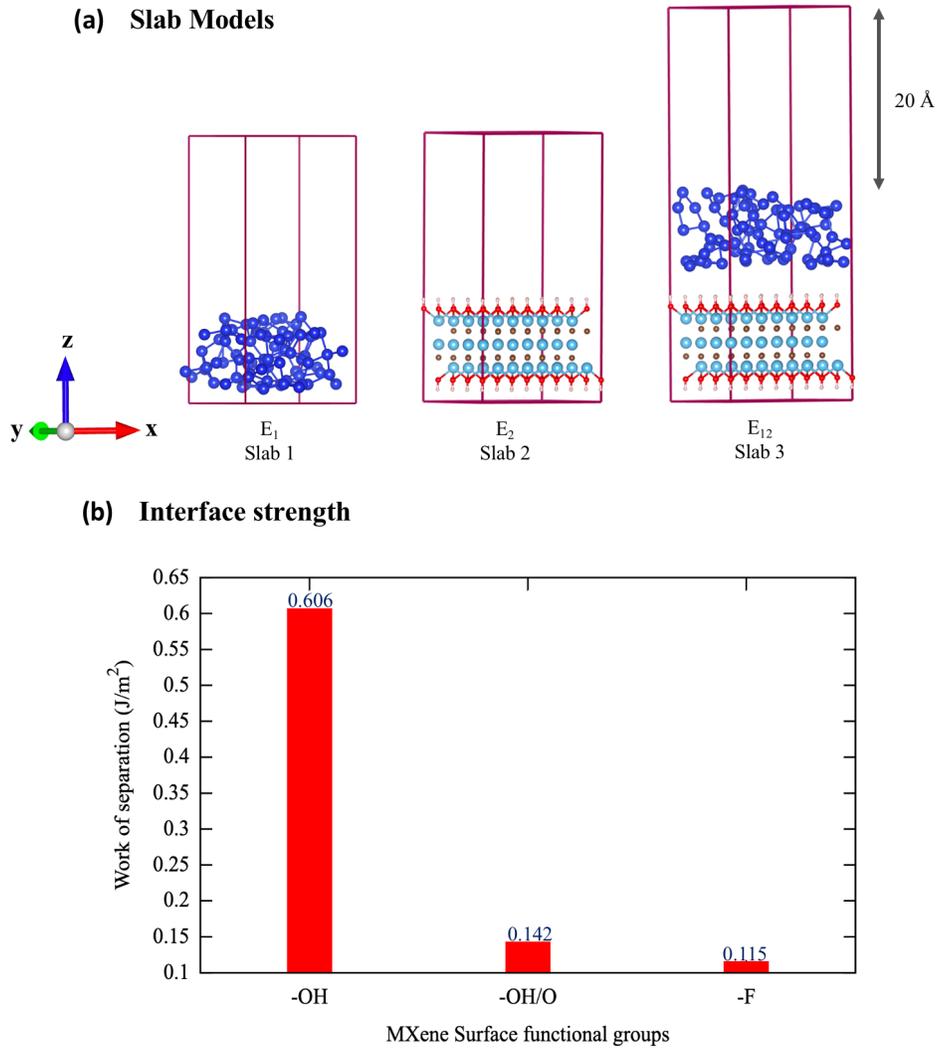

**Figure 2 Vacuum slab model for energy calculations and interface strength results. (a)** Representation of slab model used for the calculation of work of separation ($W_{sep}$) between *a*-Si bulk and MXenes. Slab 1 consists of amorphous Si, slab 2 consists of functionalized $Ti_3C_2T_x$ MXene monolayer, and slab 3 has interface system of *a*-Si over the respective MXene. **(b)** Interface strength between *a*-Si bulk and $Ti_3C_2T_x$ MXenes with changing surface functional groups ($T_x$), as calculated by $W_{sep}$.

**Table 1** Final energies, interfacial gap *d* and equilibrium dimensions of *a*-Si/$Ti_3C_2T_x$ interfaces. For each interface system, **$E_1$** is energy of slab 1, **$E_2$** is energy of slab 2, **$E_{12}$** is the total energy of interface system in slab 3, and **A** is the area of contact at the interface. Interfacial gap *d* is the vertical distance noted between lowest Si atom and top surface atoms of MXene substrate, in the optimized structure.

| S. no. | Functional Group | d (Å) | DFT optimized energy (eV) | | | Box Dimensions after optimization | | |
|---|---|---|---|---|---|---|---|---|
| | | | $E_1$ | $E_2$ | $E_{12}$ | x (Å) | y (Å) | Area (Å$^2$) |
| *(i)* | T = OH | 2.34 | -223.515 | -921.225 | -1149.701 | 12.333 | 12.344 | 131.65 |
| *(ii)* | T = OH/O | 2.26 | -223.481 | -887.087 | -1111.724 | 12.293 | 12.297 | 130.60 |
| *(iii)* | T = F | 2.14 | -223.457 | -724.417 | -948.817 | 12.286 | 12.272 | 130.81 |

*3.2 Effect of interfacial gap (d)*

As much as we advocate low interfacial strength for the smooth long-lasting working of Si/MXene electrodes, we strongly recommend interfacial strength to remain above a threshold value to prevent complete loss of electronic contact between the two materials. Studies on the interface properties of 2D materials with 3D bulk are still in their infancy. Thus quantitative determination of threshold value of interface adhesion for the continued electronic contact will require more detailed analysis with application-specific experimental validation. Since this lies beyond this study's purpose, we assume the negative values for $W_{sep}$ will be universally derogatory for all the interface systems. Thermodynamically, interface strength is sensitive to the energy of interface system $E_{12}$. In equation 1, $W_{sep}$ depends on the difference between $E_{12}$ and the sum of energies of the individual materials ($E_1 + E_2$). If $E_{12}$ is lower than $E_1 + E_2$, both materials can come together to form a stable interface with a positive $W_{sep}$, as is the case for the three interface systems presented in Table 1. In contrast, the high system energy of the interface $E_{12}$ indicates either lack of chemical interaction or the presence of interfacial strains due to local charge redistribution. Both these conditions are the ancillary outcome of interfacial gap *d*. If *d* between the two materials is too high, there is a possibility of a lack of chemical interaction. Conversely, if *d* is too low, atoms in the interface might be strained due to lattice misfit or stearic repulsions caused by the concentration of charges.

Several earlier works on 2D material such as graphene[48, 49] throw light on the interface strength variation with the interfacial gap *d*. These computational studies summarize that the adhesion of 3D bulk materials on graphene substrate first increases and then decreases as the two materials are brought closer. To realize the same relation for MXene substrates with *a*-Si bulk, we performed AIMD simulations within the DFT framework to trace the changes in interface system energy $E_{12}$ as the interfacial gap $d$ varies due to diffusion of Si atoms in the interfacial region. AIMD simulation is a rigorous tool that provides insight into the system's dynamics at a finite temperature by calculating forces for every frame from accurate electronic structure calculations. We employed relaxed structures of three different $a$-Si/ $Ti_3C_2T_x$ interface systems (each having different MXene surface functionalization, initial $d$ ~1.5 Å, and vacuum of 20 Å in z dimension) and observed the dynamics for 1000 AIMD steps. Snapshots of the starting three configurations are shown in the insets of figure 3. The two materials are very close ( $d$ < 2 Å) in the start, which causes strain on the surface functional groups, as depicted by O-H bonds' alignment in the snapshots. During the AIMD run, system energies $E_{12}$ fluctuate as the interfacial gap $d$ is altered due to movement of loosely bound interfacial Si atoms. These changes occurred during different time frames for the three interface systems. Since our primary focus lay in observing the correlation between the interfacial gap and the system's stability, we plot only the system energy $E_{12}$ for the individual interfaces corresponding to the $d$ at that specific time frame in figure 3.

The three plots clearly demonstrate a drop in the system energy $E_{12}$ with an initial increase in interfacial gap $d$, followed by a rise of $E_{12}$ as $d$ further increases. The trend is clearly in accordance with the previous graphene-based works[48] and indicates the formation of a potential well between 2 - 2.5 Å interfacial gap for all three interface systems. This drop of energy $E_{12}$ indicates stability and agrees with our assumption that the distance of 2 - 2.5 Å between the current material surfaces should facilitate interface formation. While the value of $d$ for potential wells in figure 3 are not absolute, they represent a close range where system stability could be achieved. It is apparent in figure 3a and 3b that $a$-Si/MXene interface is most stable at the interfacial gap ~2Å when MXene surface is functionalized with -OH and -O groups. In the case of fluorinated MXene, the potential well shifts slightly towards higher $d$. Determination of absolute $d$ for potential well required very precise measurements of the distance between the two surfaces. This was not possible for our current configurations, where one material is amorphous, consisting of loosely bound and non-uniformly distributed surface Si atoms. Another important observation that can be made from the presented plot is regarding the system energy at $d$ < 2Å. Upon comparing the three interface systems, the system energy at $d$ < 2Å was very high for completely hydroxylated and fluorinated MXene interfaces (as demonstrated in figures 3a and 3c). In contrast, interface system with -OH/O mixed surface functionalization appears comparatively stable at $d$ as low as 1Å. This is plausible due to bond formation between interfacial Si atoms and reactive -O groups on the MXene surface, also visible in the snapshot of the initial configuration in figure 3b. H and F atoms on the surface of hydroxylated and fluorinated MXenes are well-coordinated and not free to form covalent bonds with Si atoms. Thus, close vicinity of Si surface causes strain on H-O, O-Ti, and F-Ti bonds, resulting in very high system energies. On the other hand, in second interface system, very close proximity of Si surface does cause certain strain to H-O bonds, but some loosely bound Si atoms diffuse closer to the surface to form Si-O bonds with the surface -O groups.

The analyses presented here reveals the dependence of interface strength on the interfacial gap $d$ in $a$-Si and $Ti_3C_2T_x$ MXene systems through system energy $E_{12}$. The existence of strains, chemisorption or physisorption at the interface are closely associated with interfacial gap $d$. Our calculated interface strengths in the previous section differed significantly due to surface functional groups ($T_x$), while the interfacial gap $d$ (in Table 1) had varied only slightly. We next investigate the variation in physico-chemical attributes of the formed interfaces and their relationship with the interface strength.

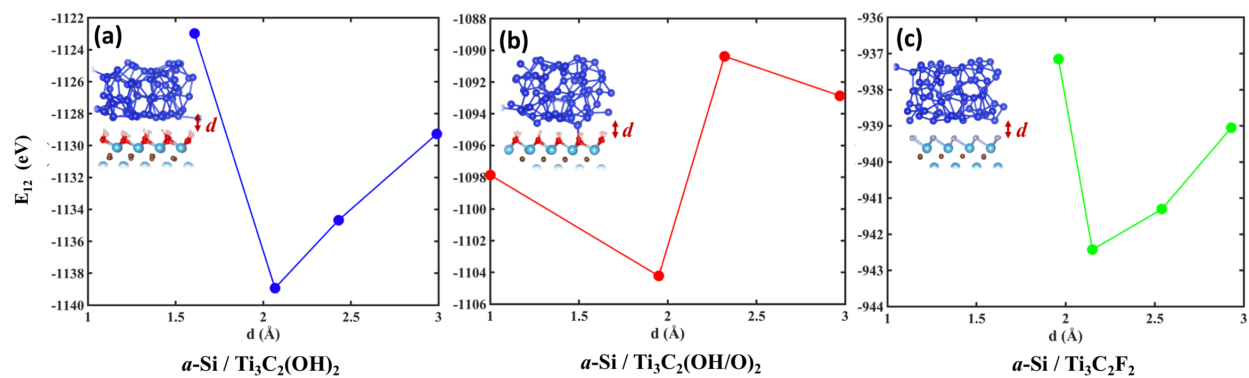

**Figure 3 Variation of interface system energy $E_{12}$ with interfacial gap $d$.** Energy profiles of interface systems as distance between the two materials expand during AIMD simulation. Inset depicts schematic representation of initial interface structure with $d$ ~1.5Å. **(a)** $a$-Si/$Ti_3C_2(OH)_2$ interface, **(b)** $a$-Si/$Ti_3C_2(OH/O)_2$ interface, and **(c)** $a$-Si/$Ti_3C_2F_2$ interface.

### 3.3 Electron distribution across the interface

To comprehend the root cause of variation in the interface strength, a complete understanding of local charge redistribution across the interface is necessary as it depends critically on the material pair. Here, we throw more light on the electron redistribution at *a*-Si/ $Ti_3C_2T_x$ interfaces as the functional groups on MXenes are varied. For this, Bader charge analysis is performed on the optimized interface systems using scripts by Henkelman group[50]. Bader charge calculation scheme quantifies atomic charges based on the charge density in each atom's bader volume in the system. Based on our used pseudopotential, Si atoms in the system have four valence electrons. Therefore, total electron transfer between the two materials (*a*-Si and $Ti_3C_2T_x$) is determined by summing electronic charges on all the Si atoms in the system. In all three interface systems, electrons were transferred from bulk *a*-Si to MXene (illustrated in figure 4 a-c) and are mentioned in Table 2 where the net charge transfer across the interface is quantified as **Δq**.

The net electron exchange (Δq) at the interface is important for two reasons: first, it is symbolic of comparative ease of electronic interaction between the two surfaces; and second, it throws light on the existing bonding phenomenon. Charge transfer across the interface systems increases as functional groups ($T_x$) change from -OH to -F in MXenes (Table 2, *i-iii*)*.* This quantitative evaluation could be explained by physico-chemical property of work function, which is the energy required to remove an electron from the surface. Yu et al.[20] earlier reported the work function of the surface functionalized MXenes as follows: -OH terminated MXene has the lowest work function of 0.44 eV, while -O and -F terminated MXenes have high work functions (6.10 eV and 4.92 eV, respectively). Consequently, $Ti_3C_2(OH)_2$ surface will have the lowest electron affinity, which will increase proportionately to the change in surface functional groups (-O and -F). Moreover, O and F atoms are highly electronegative (EN) in comparison to Si atoms ($EN_O$ = 3.44, $EN_F$ = 3.98, and $EN_{Si}$ = 1.90), and therefore, possess ability to withdraw more electrons from the latter. Thus herein, $Ti_3C_2(OH)_2$ acquire only 0.054 $e^{-1}$ from Si bulk while $Ti_3C_2F_2$ acquired the highest $e^{-1}$ count from the latter.

Conventionally, interface strength has a linear relationship with Δq which impedes as the bonding distance increases[38, 51]. Therefore, we expected $W_{sep}$ to have a linear relationship with **Δq** and **1/$d^2$** , as emphasized by former Si-C interface study[38]. Conversely, for the case of Si-MXene interfaces, a downward trend is noted between the two quantities, as illustrated in figure 4d. Mere 0.054 $e^{-1}$ are exchanged at the *a*-Si/ $Ti_3C_2(OH)_2$ interface, which has the highest interface strength among the three interface systems. On the contrary, highest Δq (2.32 $e^{-1}$) is noted for *a*-Si/ $Ti_3C_2F_2$ interface having the weakest interface strength. Thus in the case of three *a*-Si/MXene interfaces considered, interface strength cannot be assessed correctly from the quantification of Δq alone. Evaluation of bonding phenomenon and stearic effects at the interface is imperative for a thorough understanding of interface strength.

**Table 2** Summary of net electrons exchanged **(Δq)** across the interfaces along with its associated interface strength value (**W**$_{sep}$). Interfacial gap *d* here is the distance between MXene and the lowest Si atom.

| S. no. | Functional Group | Interface strength (W$_{sep}$) | Electrons exchanged (Δq) | d |
|---|---|---|---|---|
| *(i)* | T = OH | 0.606 J/m$^2$ | 0.054 e$^{-1}$ | 2.34 Å |
| *(ii)* | T = OH/O | 0.142 J/m$^2$ | 0.37 e$^{-1}$ | 2.26 Å |
| *(iii)* | T = F | 0.115 J/m$^2$ | 2.32 e$^{-1}$ | 2.14 Å |

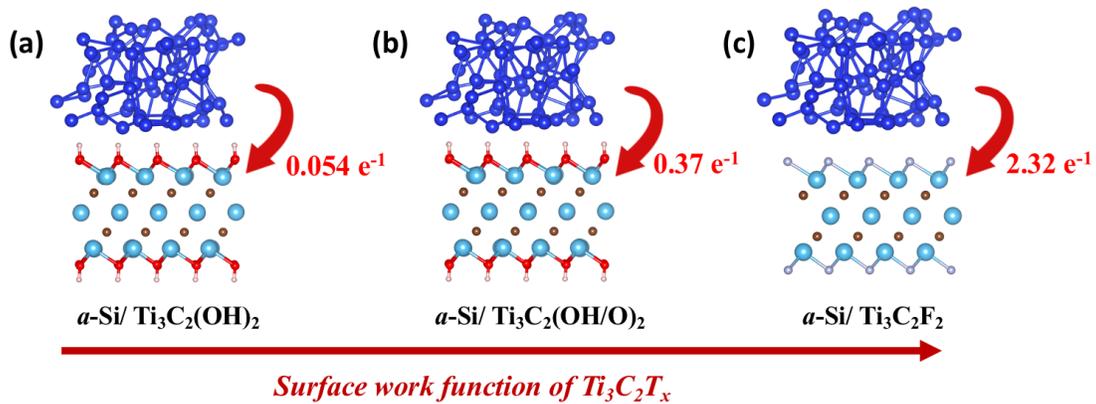

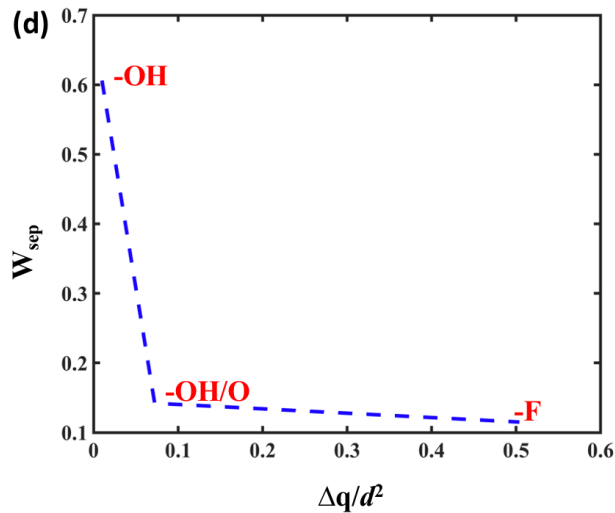

**Figure 4 Relationship between interface strength, interfacial electron exchange and surface chemistry of Mxenes. (a,b,c)** Atomic representation of a-Si/MXene interfaces depicting net charge transfer (Δq) from Si to MXene in three interface models having different surface functional groups ($T_x$). **(d)** Down trend between the calculated interface strength and the total electrons exchanged across the interfacial gap ($q/d^2$) at a-Si and MXene interface with different MXene substrates.

### *3.4 Combined effect of interfacial gap and electron distribution at the interface*

To understand the influence of surface functional groups on the charge redistribution at the atomic scale, we further zoom into atoms' the charge distribution present at the interface. Charge density in the interfacial region of *a*-Si and $Ti_3C_2T_x$ MXenes is visualized by charge separation analysis and shown in figure 5 a1-c1. Charge separation scheme at the interface was extracted by subtracting charge density of individual materials from that of the entire system, and the difference is plotted with an isosurface of 0.0007 e Å$^{-3}$. Accumulation and depletion of charges are depicted by red and green color in figure 5. We have used this analysis to throw light on the influence of interfacial gap *d* on the electron distribution and charge density customized to the atom type present at the interface. Sub figures in figure 5 (a2-c2, a3-c3), focusses on the total electrons on individual atoms (denoted by ***q***, derived by Bader charge analysis) at the interface as the interfacial conditions change (*d* and $T_x$) within a system. The sum electrons on the surface atoms of MXenes at the Si interface and the free surface are also summarized in Table 3.

Loosely bounded Si atoms in *a*-Si bulk are distributed over MXene surfaces non-uniformly. While some surface Si atoms adsorb closely on the MXene surface, the majority are at a distance > 3Å, forming weak vdW interactions with the substrate. The charge separation scheme in figure 5 a1 indicates physisorption as the primary bonding mechanism in the system, which results in an intermittent amount of interface strength (0.60 J/m$^2$). This is also favored by the lack of atomic strains on the interfacial atoms. In figure 5 a1 and b1, loss of electrons on the oxygen bound hydrogen in OH groups cause polarity in the interfacial region. In the case of T= OH , highly electronegative O extracts electrons from H and Ti atoms. This leads to the existence of nearly free electron (NFE) states parallel to the surface in the interfacial region with the highest positive charges (depicted by uniform green isosurfaces in figure 5 a1). NFE states enable electron transmission in the interfacial channel without nuclear scattering[52-54]. This makes $Ti_3C_2(OH)_2$ an ideal substrate for Si electrode particles with facilitated electron transfer. Figure 5 a2,a3 presents electron distribution on the interfacial atoms at *a*-Si/ $Ti_3C_2(OH)_2$ interface. Electron deprived H atoms can extract a small charge from a closely adsorbed Si atom (when $d = 2.34$ Å in figure 5 a2), in contrast to the high interfacial gap condition within the system when Si atom is present at the distance $d > 3.2$ Å from the MXene surface (figure 5 a3). In the latter condition, H and Si loose covalent contact and no electron exchange occurs between the two. We note in Table 3, the surface H atoms in the interfacial region have a slightly higher sum of total electrons (15.2054 $e^{-1}$) than the H atoms present on the free surface with no intimate contact with Si (14.5529 $e^{-1}$). Hence, charge depletion on H atoms directs very little electrons from the Si at the interface.

Upon replacement of few -OH groups by -O on the surface of MXene, Si atoms are noted to move away from the surface -O groups and become more localized near -OH groups. Figure 5 b1 depicts the red isosurfaces on Si atoms closer to MXene, predominantly in the region with -OH groups. There is hardly any Si atom within the bonding range of surface -O groups. These two conditions are further detailed in figure 5b2-b3. The minimum distance between Si and surface O atoms in the current system is 3.2 Å, which is not positive for forming a covalent bond between Si and O. For the possibility of Si-O bond, bonding distance should be less than 1 Å as observed in section 3.2. Consequently, the surface -O groups tend to extract more electrons from Ti atoms to stabilize (figure 5 b3). As covalency between Ti-O increases, the ability of -O to bind with Si decreases resulting in a dip of interface strength between *a*-Si and $Ti_3C_2(OH/O)_2$. Weak vdW bonds between fewer surface -OH groups and Si atoms are the only contribution to the interface adhesion. We anticipate the interface strength of such interfaces can be customized by varying the ratio of surface -OH and -O groups on MXenes. It is also noted from Table 3, the O atoms present in the interfacial region and on the free lower surface have barely any difference in the total electron count, further indicating the lack of interaction between surface -O and Si. Surface -O groups are free from Si adsorption and thus promise to enhance the electrode capacity by providing additional storage sites for Li/Na in Ion batteries[17].

At the interface of *a*-Si/ $Ti_3C_2F_2$, prominent charge density is seen around F atoms while the surrounding regions are deficient of charges( depicted by red and green isosurfaces in figure 5 c1). The interfacial gap between the two materials is lower than the previous systems ($d = 2.14$ Å), and Si atoms are seen to be more uniformly present at the interface. When the interfacial gap $d$ is as low as 2.2Å (figure 5 c2), higher electron exchange occurs between Si and F, committing to the high net electron exchange between the two materials (Table 2-*iii*, $\Delta q$=2.32 e$^{-1}$). These values indicate Si atoms are partially chemisorbed on the MXene surface. Interestingly, enhanced interaction between *a*-Si/ Ti$_3$C$_2$F$_2$ should indicate higher interface strength. However, due to close proximity of amorphous Si surface, there is a slight strain on F-Ti bonds due to stearic hindrance. When d > 3Å in the same interface system (figure 5 c3), the covalent communication between Si and F is negligent, and Ti atoms become the primary donors to F. Similar to -OH group, F takes ~ 1e$^{-1}$ from Ti atoms. Overall, the surface of -F terminated MXene becomes saturated with charges. As a consequence of high concentration of charges in the interfacial region, stearic effects between the two materials reduce the interface strength. The interface strength of *a*-Si/Ti$_3$C$_2$F$_2$ improved significantly when the interfacial gap $d$ was expanded to 3.28Å. The vdW forces hold the resultant interface with no atomic strains and $W_{sep}$ = 0.335 J/m$^2$. The two *a*-Si/ Ti$_3$C$_2$F$_2$ interface systems are compared in Supporting Information.

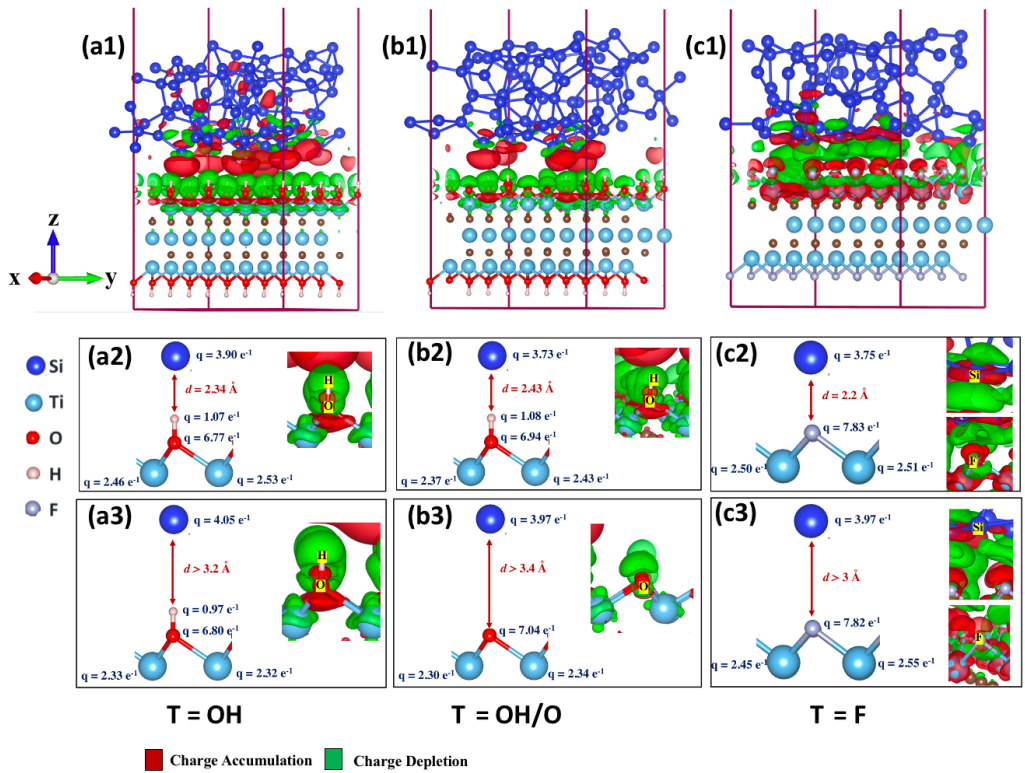

**Figure 5 Charge separation scheme and electron distribution at *a*-Si/MXene interfaces.(a1)** Charge separation scheme across *a*-Si/ $Ti_3C_2(OH)_2$ interface. **(a2-a3)** Demonstrates details of charges (q) on the atoms present at the *a*-Si/ $Ti_3C_2(OH)_2$ interface as interfacial conditions *d* change. **(b1)** Charge separation scheme across *a*-Si/ $Ti_3C_2(OH/O)_2$ interface. **(b2-b3)** Demonstrates details of charges (q) on the atoms present at the interface as the interfacial conditions d and $T_x$ change. **(c1)** Charge separation scheme across *a*-Si/ $Ti_3C_2F_2$ interface. **(c2-c3)** Demonstrates details of charges (q) on the atoms present at the interface as *d* changes.

**Table 3** Sum of total charges (q) on the MXene surface atoms present at the Si interface and at the free surface.

| Functional Group | Surface atom | Total electrons on the surface atoms in interfacial region | Total electrons on the surface atoms on lower side |
|---|---|---|---|
| T = OH | H | 15.2054 $e^{-1}$ | 14.5529 $e^{-1}$ |
| T = OH/O | H | 8.4297 $e^{-1}$ | 7.8734 $e^{-1}$ |
| T = OH/O | O | 111.4490 $e^{-1}$ | 111.1261 $e^{-1}$ |
| T = F | F | 125.4535 $e^{-1}$ | 124.53 $e^{-1}$ |

*3.5 Implications of interface analysis beyond ion batteries*

Besides ion battery electrodes and supercapacitors, there is enormous potential for MXene/Si duo in optoelectronic devices such as solar cells. The recognition of Schottky junction at the $Ti_3C_2T_x$ / Si interface with its photodetector applications[55] opened the door for their usage in solar cell applications. Lately, MXene - Si composite solar cells were reported to have enhanced efficiency of 11.5%[56]. Surface terminations inevitably impact the performance of MXene based solar cells due to variations in surface work function[57]. However, it will be crucial to take interface strength between MXene and Si into account while designing the system components. Till date, it is difficult to control the surface terminations during synthesis procedures, and most synthesized MXenes result in mixed terminations of -F/OH/O groups. Unlike in ion batteries where lower interface strength between MXene and Si is more suitable, the interface in solar cells is expected to be resistant to annealing. Thus high to intermittent interface strength is preferred for longevity and enhanced performance.

## 4. Conclusions

To conclude, we have carried out DFT calculations to quantify the variation in interface strength between 3D amorphous Si bulk and surface terminated $Ti_3C_2T_x$ MXenes. Our results show that -OH functionalized MXene binds most strongly to amorphous Si with work of separation of 0.606 J/m$^2$ in comparison to -OH/O mixed and -F functionalized MXenes. These values of interface adhesion ranged from intermittent to low and are favorable for battery applications to permit easy expansion/contraction. AIMD simulations confirm that interfacial gap $d$ between the two materials strongly influences the interface systems' energetics and stability. Generally, a potential well is noted for the interface energy when $d$ lies within 2Å and 2.5Å. But, in the case of surface fluorinated MXene, the potential well shifts slightly towards higher $d$. Next, the overall net electron exchange at the interface has little to say about the interface strength. The interface strength noted for the three interface systems is not linear to charge transferred across the interface as per the popular observation. Electron distribution across the interface is driven by physico-chemical surface properties such as work function and electronegativity of the functional groups. Detailed analyses of interfacial gap and bonding mechanism reveal that physisorbed interfaces have better interface strength as noted for *a*-Si/ $Ti_3C_2(OH)_2$ and *a*-Si/ $Ti_3C_2F_2$. The presence of a high concentration of charges in the interfacial region of partially chemisorbed materials resulted in steric effects. It was ultimately responsible for low interface strength as in the case of -F terminated $Ti_3C_2T_x$ MXene and Si. Our results provide more in-depth insight into the atomic-level interfacial phenomena of surface terminated MXene with Si.

## AUTHOR INFORMATION

**Corresponding Authors**

*Email : dibakar.datta@njit.edu

*Email : vs574@njit.edu

**Author Contributions**
D.D. and V.S. designed the project. V.S. performed all computations and wrote the manuscript. All authors have given approval to the final version of the manuscript.

**Notes**

The authors declare no competing financial interest.

**ACKNOWLEDGEMENT**

The work is supported by NSF (Award Number 1911900). D.D. acknowledges Extreme Science and Engineering Discovery Environment (XSEDE) for the computational facilities (Award Number – DMR180013).

**DATA AVAILABILITY**

The data reported in this paper is available from the corresponding author upon reasonable request.

**CODE AVAILABILITY**

The pre- and post-processing codes used in this paper are available from the corresponding author upon reasonable request. Restrictions apply to the availability of the simulation codes, which were used under license for this study.